\documentclass[titlepage,12pt]{utarticle}  
\usepackage{amssymb}
\usepackage{floatflt,graphicx} 
\usepackage[]{hyperref} 

\numberwithin{equation}{section}

\newcommand{\abs}[1]{\lvert #1\rvert}

\newcommand{\sh}{\rm sinh}

\newdimen\tableauside\tableauside=1.0ex
\newdimen\tableaurule\tableaurule=0.4pt
\newdimen\tableaustep
\def\phantomhrule#1{\hbox{\vbox to0pt{\hrule height%
	\tableaurule width#1\vss}}}
\def\phantomvrule#1{\vbox{\hbox to0pt{\vrule width%
	\tableaurule height#1\hss}}}
\def\sqr{\vbox{%
  \phantomhrule\tableaustep
\hbox{\phantomvrule\tableaustep\kern\tableaustep\phantomvrule\tableaustep}%

\hbox{\vbox{\phantomhrule\tableauside}\kern-\tableaurule}}}
\def\squares#1{\hbox{\count0=#1\noindent\loop\sqr
  \advance\count0 by-1 \ifnum\count0>0\repeat}}
\def\tableau#1{\vcenter{\offinterlineskip
  \tableaustep=\tableauside\advance\tableaustep by-\tableaurule
  \kern\normallineskip\hbox
    {\kern\normallineskip\vbox
      {\gettableau#1 0 }%
     \kern\normallineskip\kern\tableaurule}%
  \kern\normallineskip\kern\tableaurule}}
\def\gettableau#1 {\ifnum#1=0\let\next=\null\else
  \squares{#1}\let\next=\gettableau\fi\next}

\begin{document}

\preprint{
UTTG--07--99\\
{\tt hep-th/9911040}\\
}
\title{
Chiral Symmetry Breaking\\
in the AdS/CFT Correspondence
}
\author{
Jacques Distler and Frederic Zamora
  \thanks{Work supported in part by NSF Grant PHY9511632 and
  the Robert A.~Welch Foundation.}
\oneaddress{
    Theory Group, Physics Department\\
    University of Texas at Austin\\
    Austin TX 78712 USA.\\
{~}\\
\email{distler@golem.ph.utexas.edu}\\
\email{zamora@zerbina.ph.utexas.edu}}
}
\date{November 5, 1999\\ revised April 10, 2000}

\Abstract{
We study the $SU(3)$-invariant relevant deformation
of $D=4$ ${\cal N}=4$ $SU(N)$ gauge theory at large $N$
using the AdS/CFT correspondence. 
At low energies, we obtain a nonsupersymmetric gauge theory with
three left-handed quarks in the adjoint of $SU(N)$.
In terms of the five dimensional gauged supergravity, there is an 
unstable critical point in the scalar potential
for fluctuations of some fields in a nontrivial representation 
of the symmetry group $SU(3)$.
On the field theory side, this corresponds to dynamical 
breaking of the $SU(3)$ chiral symmetry down to $SO(3)$.
We compute the condensate of the quark bilinear and the two-point
correlation function of the spontaneously broken currents from supergravity
and find a nonzero `pion' decay constant,
$f_\pi$.
}

\maketitle


\section{Introduction}\label{sec:intro}

Consider $SU(N)$ Yang-Mills fields, coupled to $N_{f}$ Weyl fermions,
$\lambda^i$ ($i=1,...,N_f$),
in the adjoint of the gauge group $SU(N)$, often referred to as `adjoint QCD'.
For $N_{f}\leq5$, the theory is asymptotically
free and very plausibly confines for energies
smaller than the dynamically generated scale.
The global symmetry group
is $SU(N_f)$. The  Weyl fermions are in
the fundamental representation, which is complex for $N_f>2$.
This chiral symmetry is presumably dynamically broken down to $SO(N_{f})$,
its largest vector-like subgroup%
\footnote{Actually, the chiral symmetry group is
    $(SU(N_{f})\times\BZ_{2N_{c}N_{f}})/\BZ_{N_{f}}$,
    broken to $O(N_{f})$; but, since we are not worried about things
    like domain walls, it suffices to consider a connected component of
    the vacuum manifold.
}
\cite{Coleman:1980mx,Vafa:1984tf}.
The order parameter for chiral symmetry breaking
is the expectation value of the quark bilinear
\begin{equation}
    \langle {\rm tr}\lambda^i\lambda^j \rangle\propto \delta^{ij}
    \quad\quad i,j =1,...,N_f\quad.
\label{eq:bilinear}
\end{equation}

In the infrared, the physical degrees of freedom
are $N_f(N_f+1)/2-1$ massless Goldstone bosons,
whose dynamics is described by
an effective chiral Lagrangian -- a nonlinear $\sigma$-model, whose
target space is $SU(N_{f})/SO(N_{f})$.
At least for small
Goldstone boson fields, we can parametrize the coset as
\begin{equation}
     U(x)=e^{i\pi^{a}(x)s_{a}/f_{\pi}}
     \label{eq:pion}
\end{equation}
where the $s_{a}$ are real, traceless, symmetric matrices, satisfying
$tr(s_{a}s_{b})=2\delta_{ab}$.
The most relevant term in the action contains two derivatives
and may be written as
\begin{equation}
     S_2[U]=-\frac{f_{\pi}^{2}}{4}\int d^{4}x \ {\rm Tr}
     \Bigl(\tfrac{1}{2}\bigl(U \partial
     U^{-1}+(U\partial U^{-1})^{T}\bigr)\Bigr)^{2}\quad.
     \label{eq:chiralLagrangian_2deriv}
\end{equation}
One can easily check that it is invariant under global
right-$SU(N_{f})$ and \emph{local} left-$SO(N_{f})$ transformations
\begin{equation}
     U(x)\to h(x) U(x) g^{-1},\quad g\in SU(N_{f}),\quad h(x)\in SO(N_{f})\quad.
     \label{eq:transformation}
\end{equation}
The dimensionful constant $f_\pi$ is related to the strength of
  the chiral fermion condensate \eqref{eq:bilinear}.

There is a one-loop chiral anomaly associated to
the $SU(N_f)$ symmetry group, whose manifestation
at low energies is encoded by the Wess-Zumino action%
\footnote{Here and below, we work in Euclidean signature.}.
Using some extension of $U(x)$ to the 5-ball, the WZ action is
\begin{equation}
     S_{WZ}[U]=-\frac{i(N^2-1)}{120\pi^{2}} \int_{B_{5}} {\rm Tr}
     \Bigl(\tfrac{1}{2}\bigl(
     U d U^{-1}+(U d U^{-1})^{T}\bigr)\Bigr)^{5}
     \label{eq:SWZ}
\end{equation}
and is the leading term odd in the number of Goldstone boson fields.
The normalization is \emph{twice} what we naively expect for $U(x)\in SU(N_f)$
\footnote{The quantization of the WZ action is slightly subtle here. As we will
    mainly be interested in $N_f=3$, let us specialize to that case (similar
results
    hold for $N_f=4,5$).
    The result \eqref{eq:SWZ} is twice what we would
    get by normalizing it to the integral generator of  $\coho{5}{SU(3)}$.
    Consider the fiber bundle $SU(3)\stackrel{\pi}{\to}SU(3)/SO(3)$,
    with fiber $SO(3)$. We claim that if $y$ is the generator of
    $\coho{5}{SU(3)/SO(3),\BZ}$, then $\pi^{*}y$ is {\it twice} the
    generator of $\coho{5}{SU(3),\BZ}$. We relegate the details to Appendix B.
}.
It is invariant under global
right-$SU(N_{f})$ and \emph{topologically-trivial} local left-$SO(N_{f})$
transformations. However  $\pi_4(SO(N_{f})\neq0$. The Wess-Zumino action shifts
by $2\pi i$ under homotopically-nontrivial $SO(N_{f})$ transformations. Were it
not
for the extra factor of 2 in \eqref{eq:SWZ}, $e^{-S_{WZ}[U]}$ would change sign
under $SO(N_{f})$ transformations which are nontrivial elements of $\pi_4$.

We are interested in correlation functions of
the $SU(N_f)$ conserved currents, $J$. At low momenta (or widely separated
points), these may be computed from the
Goldstone boson effective Lagrangian.
The simplest way to study such correlators is to
add sources $A$ for the currents $J$ to the two-derivative
action \eqref{eq:chiralLagrangian_2deriv},
\begin{equation}
     S_2[U,A]=- \frac{f_{\pi}^{2}}{4}\int d^{4}x\ {\rm Tr}
     \Bigl(\tfrac{1}{2}\bigl(U (\partial +A)
     U^{-1}+(U(\partial+A)
U^{-1})^{T}\bigr)\Bigr)^{2}
\label{eq:gaugedchiralLagrangian}
\end{equation}
in a way that now \eqref{eq:gaugedchiralLagrangian}
is invariant under \emph{local} $SU(N_f)$ transformations, given by
\eqref{eq:transformation} and
\begin{equation*}
     A\to g A g^{-1}+g d g^{-1}\quad.
\end{equation*}
Similarly, one can gauge the Wess-Zumino term, $S_{WZ}[U,A]$,
such that its variation under local $SU(N_f)$ transformations
reproduces the chiral anomalies of the underlying fermion theory
(for a discussion of the general case of $G/H$, see \cite{Zumino:G/H}).

By differentiating with respect to $A$ and setting $A=0$, we obtain
the correlation functions of the currents.
In momentum space, the two point function is
\begin{equation}
     \langle
J_{\mu}^{a}(p_{1})J_{\nu}^{b}(p_{2})\rangle\underset{p\to0}{=}
f_{\pi}^{2}
     \delta^{ab}\frac{p_{\mu}p_{\nu}}{p^{2}}
     (2\pi)^{4}\delta^{(4)}(p_{1}+p_{2})\quad.
     \label{eq:two-point}
\end{equation}
The three point function contains the anomaly
\begin{equation}
     p_{1}^{\mu}\langle J^{a}_{\mu}(p_{1}) J^{b}_{\nu}(p_{2})
         J^{c}_{\lambda}(p_{3})\rangle
	=-\frac{i}{2^8 3\pi^{2}}(N_{c}^{2}-1)d^{abc}
	\epsilon^{\nu\lambda\rho\sigma}p_{2\rho}p_{3\sigma}
	(2\pi)^{4}\delta^{(4)}(p_{1}+p_{2}+p_{3})
\label{eq:three-point}
\end{equation}
where $d^{abc}=\tfrac{1}{2}Tr(s^{a}\{s^{b},s^{c}\})$. So by studying
the two- and three-point functions of the currents at low momenta, we
can determine $f_\pi$.

For $N_f=3$, this theory can be embedded in ${\cal N}=4$ SYM as
follows. Starting with the ${\cal N}=4$ theory, we add a mass term
for one of the gluinos,
\begin{equation}
\Delta L = {\rm tr}( m \lambda^4\lambda^4 +
\overline{m}\overline{\lambda}_4\overline{\lambda}_4)\quad.
\label{eq:masspert}
\end{equation}
  This breaks all the supersymmetries and breaks $SU(4)_{R}$ to $SU(3)$.
Integrating out the massive fermion at one loop (exact, since the
fermions appear only quadratically in the action), we obtain a common
mass-squared for all six scalars in the ${\cal N}=4$ vector multiplet.
To see this, it is convenient to write the ${\bf 6}$ as the
antisymmetric product of two ${\bf 4}$s. A convenient basis,
$\phi_{ab}$ ($a,b=1,...4$), for the scalars is ($I=1,2,3$)
\begin{equation*}
     \begin{split}
         \phi_{i4}&=\Phi_{I}+i\Phi_{I+3}\\
         \tfrac{1}{2}\epsilon^{ijk}\phi_{jk}&=\Phi_{I}-i\Phi_{I+3}\quad.
     \end{split}
\end{equation*}
The Yukawa couplings are
\begin{equation*}
      \sqrt{2}\ {\rm tr}\bigl( \phi_{ab}(\lambda^{a}\lambda^{b}
+\epsilon^{abcd}\overline{\lambda}_{c}\overline{\lambda_{d}})\bigr)\quad.
\end{equation*}
The one-loop diagram with two insertions of the fermion mass
matrix, $m_{ab}$, induces a scalar mass term proportional to
\begin{equation*}
     \raisebox{-6.5ex}{\includegraphics[width=36ex]{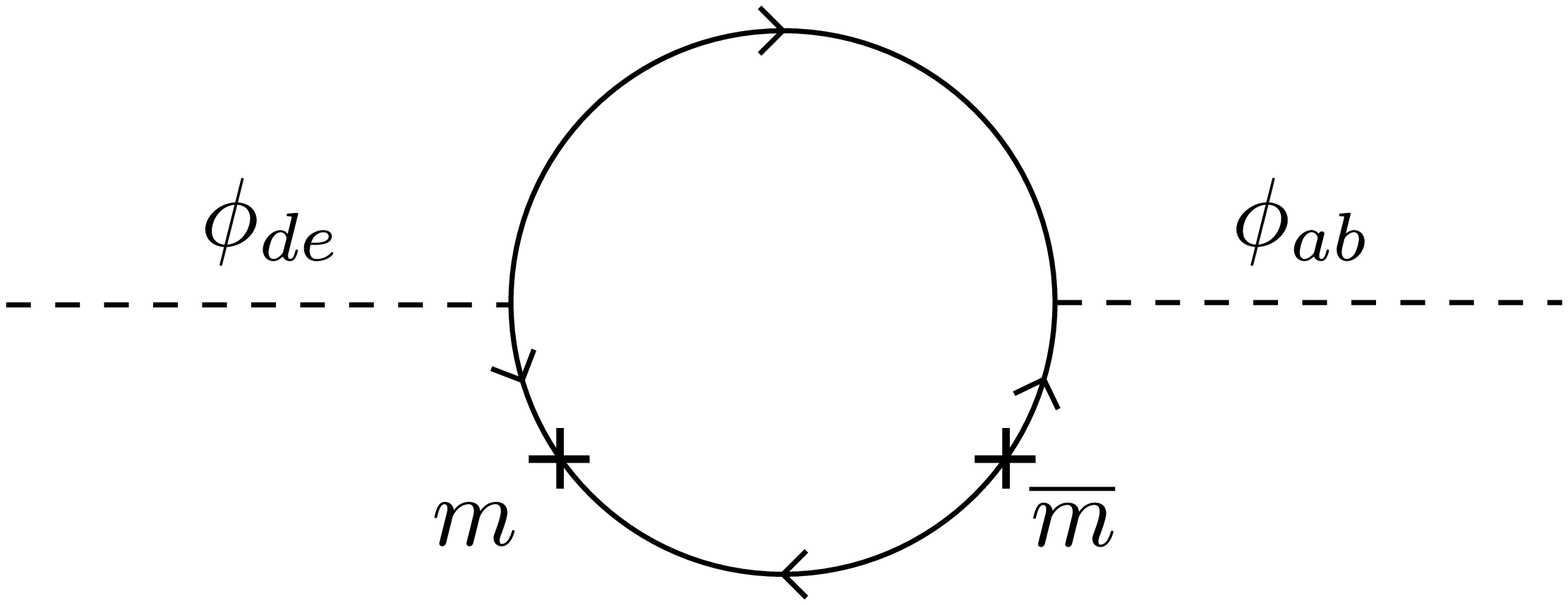}}
     \sim\epsilon^{acde}(m\overline{m})_{c}{}^{b}\phi_{ab}\phi_{de}
\end{equation*}
which, for $m_{44}=m$ and otherwise zero,
is just a common mass for all six scalars.

Thus, at low energies, we obtain a non-supersymmetric $SU(N)$
gauge theory with three massless adjoint
fermions and no scalars -- in other words, adjoint QCD for
$N_{f}=3$. We use the ${\cal N}=4$ theory as a
particular ultraviolet regularization of this $N_{f}=3$ theory.
For large $N$,
we can try to exploit the AdS/CFT correspondence to study it.

In \cite{Distler:1998gb,Girardello:1998pd},
it was observed that there is a supergravity solution
which interpolates between
the maximally supersymmetric $SU(4)$-invariant
critical point of the 5D supergravity potential, located
at the origin of the scalar field space, and the
non-supersymmetric $SU(3)$-invariant critical point
located at some point
$\sigma = \sigma_\star \not=0$ on the $SU(3)$ invariant
direction, parametrized by $\sigma$, of the scalar field space.
Both critical points have negative cosmological constant
and correspond to anti-de Sitter spaces \cite{Gunaydin:1986cu}.

In light of the AdS/CFT correspondence
\cite{Maldacena:1997re,Gubser:1998bc,Witten:1998qj},
the conformal field associated to the supergravity field $\sigma$ is
the fermion mass term \eqref{eq:masspert}
and references \cite{Distler:1998gb,Girardello:1998pd}
interpreted the interpolating supergravity solution from
$\sigma=0$ to $\sigma_\star \neq 0$ as the renormalization group flow
of the ${\cal N}=4$ $SU(N)$ gauge theory,
at large $N$ and strong 't Hooft coupling, perturbed by the
relevant perturbation \eqref{eq:masspert}.

If we believe in this
supergravity solution, we would be led to expect that the boundary
field theory flows from the ${\cal N}=4$ superconformal fixed point
in the UV to a nonsupersymmetric conformal fixed point in the IR.
This is in sharp contrast to the field theory expectations
outlined above. Admittedly,
the supergravity description is valid at large $N$ and strong 't~Hooft
coupling, where our grasp of the field theory is less secure. But
still, a nontrivial infrared fixed point for this theory would be a
surprise.

Anti-de Sitter spacetimes with scalar fields have
to satisfy a particular lower bound for the masses of such scalars,
the Breitenlohner-Freedman bound \cite{ Breitenlohner:GaugedSUGRAstability}
(see also \cite{Mezincescu:1985ev}), in order
to have an (at least perturbatively) stable gravitational background
under scalar fluctuations.
Hence, in \cite{Distler:1998gb} we stressed the importance
of performing an explicit check of the  Breitenlohner-Freedman bound
in non-supersymmetric gravitational backgrounds%
\footnote{If the background preserves some
supersymmetries, the Breitenlohner-Freedman bound is guaranteed.}.
Among the models we discussed in \cite{Distler:1998gb} was
an $SU(3)$-invariant five dimensional anti-de Sitter
background, which we found to be stable.

We now believe that calculation was in error, and the  $SU(3)$ critical point
is unstable%
\footnote{We would like to thank Freedman,
    Gubser, Pilch and Warner for pointing out to us that the stability
    analysis of the $SU(3)$ point in \cite{Distler:1998gb}
was probably incorrect.
    We are especially grateful to Krzysztof Pilch for a lengthy series of
    email exchanges in which we reconciled our respective calculations.}.
In the five dimensional ${\cal N}=8$ supergravity multiplet
there are ten complex scalars in the irrep ${\bf 10}$ of $SU(4)$.
Decomposing this representation under its
$SU(3)$ subgroup, ${\bf 10}={\bf 1}+{\bf 3}+{\bf 6}$, we obtain
the singlet, $\sigma$, a triplet which is higgsed for $\sigma\not=0$
and six physical complex scalars in
the symmetric tensor, $\tableau{2}$, of $SU(3)$.
Upon correcting our previous calculation,
it turns out that the scalars in the ${\bf 6}$ are unstable modes, which
violate
the Breitenlohner-Freedman bound%
\footnote{In Appendix A we write the scalar spectrum at the
$SU(3)$ critical point}.
The tiniest fluctuations drive us away
{}from the would-be $SU(3)$ critical point.

What does this mean for the field theory?
If the $SU(3)$ critical point is unstable, the renormalization
group flow of the $SU(3)$ invariant theory certainly
does not end up at any non-trivial infra-red fixed point.
But then, where does it flow to?
To elucidate this question, first we should identify the
conformal fields in the boundary theory which are associated to the ${\bf 6}$.
They are the fermion bilinears
\begin{equation*}
{\cal O}^{ij}={\rm tr}\lambda^i\lambda^j\quad.
\end{equation*}
{}From our previous discussion of chiral symmetry
breaking in adjoint QCD, the most plausible interpretation is that,
even though we do not couple ${\cal O}^{ij}$ to a source at the boundary
{}-- as we do for the conformal field which couples to $\sigma$ --
the renormalization group flow, read on the supergravity field space,
is driven in a particular direction where some unstable modes
in the ${\bf 6}$ are turned on. That is, whereas the $SU(4)$ symmetry
is broken explicitly, the $SU(3)$ symmetry is broken \emph{dynamically} by an
  expectation value for the operator ${\cal O}^{ij}$.

The purpose of this paper is to examine these phenomena from the
point of view of the five dimensional supergravity and,
in particular, to elucidate
the fate of the unstable $SU(3)$ critical point.
One of the main goals is to search for Goldstone boson
effects from the supergravity side. We will compute the
`pion' decay constant $f_\pi$ in this model of adjoint QCD from
the two point correlation function of the spontaneously broken
currents $J^\mu$.

\section{The Supergravity Background}

Under the $SO(3)$ subgroup of $SU(3)$ that leaves invariant
the condensate \eqref{eq:bilinear}, irreps of $SU(3)$ have the following
branching rules:
\begin{equation}
\begin{split}
{\bf 3} &= {\bf 3}
\\
{\bf 6} &= {\bf 5 + 1}
\\
{\bf 8} &= {\bf 5 + 3}\quad.
\end{split}
\end{equation}
The supergravity field corresponding to the chiral condensate
\eqref{eq:bilinear} is the $SO(3)$ singlet
in the ${\bf 6}$ of $SU(3)$. We parametrize this complex scalar as
$\chi e^{i\beta}$.
Besides this field, there are two additional
complex scalars which are singlets of $SO(3)$
in the 5D supergravity multiplet
with nonzero  boundary conditions.
There is the $SU(4)$-invariant dilaton and axion,
parametrized by the complex scalar $\rho e^{i\alpha}$,
and the $SU(3)$-invariant complex field $\sigma e^{i\gamma}$,
whose source term is the nonzero gluino mass $m$
\footnote{There is another $SO(3)$ singlet in the
    ${\bf 6}$ inside the ${\bf 20'}$ of $SU(4)$.
    But only the ${\bf 6}$ in the ${\bf 10}$ are
    unstable at the $SU(3)$ critical point.
    Hence, in order to simplify our discussion, we will
    put that additional $SO(3)$ singlet to zero.
}.

\subsection{Supergravity Lagrangian and Equations of Motion}

The next thing we need is the exact dependence of the ${\cal N}=8$
five-dimensional supergravity Lagrangian on these $SO(3)$-singlet
scalars. Details of its parametrization and of the computation
are contained in Appendix A
\footnote{When this paper was being written, the preprint
    \cite{Girardello:1999bd} appeared, where the supergravity
    Lagrangian for the fields $\sigma$ and $\chi$ (but not the dilaton or the
phases)
    is also computed.
    Interestingly, they look for BPS solutions in supergravity
    where the role of the
   boundary conditions for $\sigma$ and $\chi$ are interchanged with
    respect to ours. This simple switch gives completely different
    physics for the boundary field theory.
}.
Here we just present the result:
\begin{equation}
\begin{split}
&{\cal L} = \frac{\sqrt{g}}{2\kappa_5^2}
\left\{-\frac{1}{4}R
+\frac{1}{2}(\partial\chi)^2 +\frac{1}{2}(\partial\sigma)^2 \right.
\\
&\left.+\frac{1}{4}\left(1+\cosh^3(
\frac{2\chi}{\sqrt{3}})\cosh(2\sigma)
-\cos(2\alpha+3\beta+\gamma)
\sinh^3(\frac{2\chi}{\sqrt{3}})\sinh(2\sigma)\right)\right.
\\
&\cdot\left((\partial\rho)^2 +\frac{1}{4}\sh^2(2\rho)
(\partial\alpha)^2 \right)
\\
&\left.+\frac{1}{2}\sin(2\alpha+3\beta+\gamma)
\sinh^3(\frac{2\chi}{\sqrt{3}})
\sinh(2\sigma)\sinh(2\rho)\,(\partial\rho)(\partial \alpha) \right.
\\
&\left.+\frac{3}{8}\sinh^2(\frac{2\chi}{\sqrt{3}})
\left(\partial\beta -\sinh^2(\rho)\,\partial \alpha\right)^2
+\frac{1}{8}\sinh^2(2\sigma)
\left(\partial\gamma +\sinh^2(\rho)\,\partial \alpha\right)^2
+ V(\chi,\sigma)\right\}\quad.
\label{eq:Lagr}
\end{split}
\end{equation}
The potential only depends on the magnitudes
$\chi$ and $\sigma$:
\begin{equation}
V(\chi,\sigma) = -\frac{3}{16 L^2} \left\{
8 + \cosh(\frac{4\chi}{\sqrt{3}})-\cosh(4\sigma)
+4\cosh(\frac{2\chi}{\sqrt{3}} -2\sigma)
+4 \cosh(\frac{2\chi}{\sqrt{3}} +2\sigma) \right\}\quad.
\label{eq:pot}
\end{equation}

If $2\alpha+3\beta+\gamma =0$, one can truncate the theory to
the space of solutions with constant phases $\alpha,\beta,\gamma =0$.
This is not an accident. The effective CP-violating theta angle of the
boundary theory
turns out to be proportional to the combination of angles
$2\alpha+3\beta+\gamma$. If it is non zero, the strong CP-violating
phase starts to run under the renormalization group flow,
with the flow triggered by the supergravity equations of motion.
In order to focus on the mechanism of chiral symmetry
breaking from supergravity in the simplest possible context,
we will work in the sector of vanishing theta angle.
In this sector, the equations of motion involve only the  magnitudes
$\chi,\sigma,\rho$. Observe that the kinetic term
of $\rho$ is not canonical.
For vanishing phases, it has the scalar-dependent coefficient
\begin{equation}
\label{eq:Krho}
{\cal K}_{\rho\rho}(\chi,\sigma) =
\frac{1}{4}\left\{\cosh^2(\sqrt{3}\chi-\sigma)
+3\cosh^2(\frac{\chi}{\sqrt{3}}+\sigma) \right\}\quad.
\end{equation}

Let us now write the supergravity equations of motion
derived from \eqref{eq:Lagr}.
We parametrize the five-dimensional metric%
\footnote{We'll always work in the Einstein frame,
and in Euclidean signature.}
by the Poincar\'e-invariant ansatz
\begin{equation}
ds^2 = e^{-2y(z)} \left( dz^2 + d{\bf x}^2 \right)\quad.
\end{equation}
For AdS$_5$ with radius $L$, we have $y(z) = \log(z/L)$, with
the boundary located at $z=0$.
Then the equations of motion read:
\begin{subequations}
\label{eq:zeqs}
\begin{align}
\frac{d}{dz}\left(e^{-3y}{\cal K}_{\rho\rho}\,\dot\rho \right) &= 0
\label{eq:Eqdz}
\\
\ddot\chi -3\,\dot{y}\,\dot\chi
-\frac{1}{2}(\partial_\chi {\cal K}_{\rho\rho})\,\dot\rho^2
- e^{-2y}(\partial_\chi V) &=0
\label{eq:Eqcz}
\\
\ddot\sigma -3\,\dot{y}\,\dot\sigma
-\frac{1}{2}(\partial_\sigma {\cal K}_{\rho\rho})\,\dot\rho^2
-e^{-2y}(\partial_\sigma V) &=0
\label{eq:Eqsz}
\\
6\,\dot{y}^2 -\dot\chi^2 -\dot\sigma^2 -{\cal
K}_{\rho\rho}\,\dot\rho^2
  +2\,e^{-2y}\,V &= 0
\label{eq:E1z}
\\
2\,\ddot{y} -\dot\chi^2 -\dot\sigma^2 -{\cal K}_{\rho\rho}\,\dot\rho^2
-\frac{2}{3}\,e^{-2y}\,V &=0\quad.
\label{eq:E2z}
\end{align}
\end{subequations}
The last two equations are the Einstein's equations. It is easy
to verify that \eqref{eq:E2z} is a consequence of the previous
equations.

This coupled system of nonlinear second order differential
equations, plus the boundary conditions written below,
codify the renormalization group flow of the
${\cal N}=4$ $SU(N)$ theory deformed by the
$SU(3)$-invariant relevant coupling $m$,
and the {\it possible} chiral symmetry breaking pattern $SU(3)\to
SO(3)$.

\subsection{Ultraviolet Behavior}

The ultraviolet theory is determined by the behavior of
the supergravity solution for $z\ll L$.
The boundary conditions for $z\to 0$ determine the Lagrangian.
In our case they are
\begin{subequations}
\begin{align}
&y\to \ln(\frac{z}{L})
\\
&\rho \to \rho_0
\\
&\sigma \to mz
\\
&\chi \to 0
\label{eq:cboundary}
\end{align}
\end{subequations}
where by \eqref{eq:cboundary} we mean that $\chi$ goes to zero
faster than $z$.
One can easily find the perturbative solution satisfying these
boundary conditions:
\begin{subequations}
\label{eq:uvsol}
\begin{align}
y(z) &= \ln (\frac{z}{L}) \left( 1 +\frac{8}{15} m^4 z^4
+{\cal O}(z^6)\right) +\frac{1}{9}m^2 z^2
-\frac{m}{60}\left(\frac{524 m^3}{135}-12 \Lambda_\sigma^3 \right) z^4
+{\cal O}(z^6)
\\
\sigma(z) &= m z \left( 1 +\frac{8}{3}m^2 z^2 \ln (\frac{z}{L})
+\frac{\Lambda_\sigma^3}{m} z^2 +\frac{274}{45}m^4 z^4 \ln
(\frac{z}{L})
+{\cal O}(z^4) \right)
\\
\chi(z) &= \Lambda_\chi^3 z^3 \left( 1
-\frac{1}{6}m^2 z^2 +\frac{2}{45}m^4 z^4 \ln (\frac{z}{L})
+\frac{m}{300}\left(5\Lambda_\sigma^3 +14 m^3\right)z^4 +{\cal
O}(z^6)\right)
\\
\rho(z) &= \rho_0 +\frac{\Lambda_\rho}{L^3} z^4 \left(\frac{1}{4}
-\frac{1}{9}m^2 z^2 +{\cal O}(z^4) \right)\quad.
\end{align}
\end{subequations}

Observe the appearance of the mass scales $\Lambda_\chi, \Lambda_\rho$
and $\Lambda_\sigma$. They arise because, for a given scalar with
mass $M^2$ in anti-de Sitter space, there are two
possible roots for the scaling dimension of the boundary conformal
field to which it couples,
\begin{equation}
\Delta_\pm = 2 \pm \sqrt{4 +M^2 L^2}
\end{equation}
For the ${\cal N}=4$ fixed point, the appropriate branch is
$\Delta_+$. As we will see, the scales $\Lambda$, involving
the $\Delta_-$ root, are related to the condensates of the
corresponding operators\footnote{As noted in \cite{Freedman:1999gp},
the correspondence between the branch $\Delta_-$
and the condensate has to be refined when
the branch $\Delta_+$ is also present in the
supergravity solution for a given bulk field.}
\cite{Balasubramanian:1998de,Klebanov:1999tb}.

\subsection{Infrared Behavior}

Since the supergravity potential \eqref{eq:pot} does not have
critical points when $\chi \not=0$,
the equations of motion (\ref{eq:zeqs}) drive
the fields $\chi$ and $\sigma$ to infinity.
This divergence usually produces a naked singularity,
located at $z=z_0$, where the metric factor
$e^{-2y}$ vanishes%
\footnote{Some references where naked singularities also appear
in the study of supergravity duals are
\cite{Nojiri:1998yx,
Minahan:1999yr,Kehagias:1999tr,Gubser:1999pk,Girardello:1999hj,
Kehagias:1999iy,Constable:1999ch}.}.
This indicates that the field theory flows to a free
phase in the infrared.

Analysis of the equations of motion \eqref{eq:zeqs} near
the singularity $z_0$ shows that there is a {\it unique}
solution with the requisite behavior as $z\to z_{0}^-$.
It is
\begin{subequations}
\label{eq:irsol}
\begin{align}
&y \simeq -\frac{3}{2}\log\left(\frac{z_0-z}{L}\right)
+\frac{3}{4}\log 42
-\frac{203}{4430}\left(\frac{z_0-z}{L}\right)^2 + \cdots
\\
&\chi \simeq {\sqrt
3}\,\left(-\frac{5}{4}\log\left(\frac{z_0-z}{L}\right)
+\frac{5}{8}\log 42 -\frac{167}{8860}\left(\frac{z_0-z}{L}\right)^2
+\cdots \right)
\\
&\sigma \simeq -\frac{1}{4}\log\left(\frac{z_0-z}{L}\right)
+\frac{1}{8}\log 42 -\frac{75}{1772}\left(\frac{z_0-z}{L}\right)^2
+\cdots
\\
&\rho \simeq \rho_0 +
2^{-\frac{1}{4}}\,3^{-\frac{3}{4}}\,7^{-\frac{1}{4}}
\left(\frac{z_0-z}{L}\right)^\frac{7}{2}
\left( \frac{1}{147}
-\frac{23}{48730}\left(\frac{z_0-z}{L}\right)^2 +\cdots \right)\quad.
\label{eq:dilir}
\end{align}
\end{subequations}

If the $SO(3)$-symmetry is broken explicitly,
there is also a continuous \emph{family} of ultraviolet solutions,
parametrized by
the dimensionless ratios $m_\chi/m \neq 0$ and $\Lambda_\rho L$.
Correspondingly, one expects
a continous family of infrared solutions.
There is, indeed such a continuous
family of infrared solutions.
But, in addition to the continuous family, there
is a single \emph{isolated} solution, which is what we present
in \eqref{eq:irsol}.
Because it is isolated, we expect that it corresponds to the
spontaneously broken case.

In the ultraviolet, our solution \eqref{eq:uvsol} had three
undetermined dimensionless parameters,
  $\Lambda_{\chi,\rho,\sigma}/m$.
The appearance of such undetermined parameters in the asymptotic
solution near the boundary ($z=0$) is typical; their
values are determined by the behavior of the solution in the infrared.

That is what happens here. In principle, we could integrate
our solution \eqref{eq:irsol} numerically in the region
$[0,z_0]$ and thereby determine the coefficients in \eqref{eq:uvsol}.
The remaining dimensionful parameter, $m$, corresponds to the
location, $z_{0}$, of the singularity, which is determined
by the integral of \eqref{eq:E1z}.

We do not expect any further parameters, beyond $N$,
$g_{YM}^2 N$
and $m$, describing the field theory. So it is comforting that the
complete
supergravity solution does not have any. However, since we will not
perform the numerical integration of the supergravity
equations, our predictions for the field theory behaviour will not be
as powerful as they could be.

Since the supergravity potential is independent
of the five-dimensional dilaton $\rho$, the dimensionful scale
$\Lambda_\rho$ enters as an integration constant.
The infrared solution \eqref{eq:irsol}
determines its value to be
\begin{equation}\label{eq:LambdaRhoDef}
\Lambda_\rho = \frac{1}{16\sqrt{3}L}\quad.
\end{equation}
We cannot,  in similar fashion,  give the actual values of
$\Lambda_\chi$ and $\Lambda_\sigma$  without the
explicit connection between the IR and the UV
solution.
But the fact that the IR solution has $\chi \not=0$ {\it necessarily
implies}
that $\Lambda_\chi \not=0$. We suspect that $\Lambda_\sigma=0$.

\subsection{The Ten Dimensional Geometry}

In \cite{Khavaev:1998fb},
a conjecture for the full ten dimensional metric was given.
If we parametrize the compact five dimensional internal metric by
$\{\theta^\alpha \}$ ($\alpha=1,...,5$), the full metric is
\begin{equation}
ds_{10}^2 = \Delta^{-\frac{2}{3}} ds_5^2
+\tilde{g}_{\alpha\beta} d\theta^\alpha d\theta^\beta
\end{equation}
where $\tilde{g}_{\alpha\beta}$ is
the five dimensional internal metric. It is obtained from
\begin{equation}
\Delta^{-\frac{2}{3}}\tilde{g}^{\alpha\beta} =
c K^\alpha{}_{IJ} K^\beta{}_{KL} (G^{-1})^{IJ,KL}
\end{equation}
where $\Delta = \sqrt{{\rm det}\tilde{g}_{\alpha\gamma}
\tilde{g}_0{}^{\gamma\beta}}$,
with $\tilde{g}_{0,\alpha\beta}$ the metric on the round $S^5$;
$K^\alpha{}_{IJ}=-K^\alpha{}_{JI}$ are the $15$ Killing vectors on
the $S^5$; $G^{-1}$ is the inverse matrix of the $Sp(4)$ invariant
matrix defined in (\ref{sp4inv}); and $c$ is a normalization constant.

We have computed the ten-dimensional metric
in the $SO(3)$-invariant background \eqref{eq:irsol}. The result
is complicated, and fairly unenlightening. What we do find, however, is that
  the full type IIB geometry
\emph{is} singular at $z =z_0$.

\section{Condensates and Goldstone Bosons}

\subsection{Condensates}

As was already observed in \cite{Balasubramanian:1998de,
Klebanov:1999tb},
the dimensionful scales appearing in \eqref{eq:uvsol}
are related to the condensates of the corresponding
field theory operators.
Let us first show this for the scale, $\Lambda_\rho$,
associated to the dilaton.
Take the dilaton term in the on-shell supergravity action
and vary it with respect to the
value of the background dilaton at $z=\epsilon$,
\footnote{As usual, during the computation we need to regulate the
theory by locating the boundary at $z=\epsilon$. Also, since the
supergravity action is evaluated on-shell, only boundary
terms contribute.}
\begin{equation}
    \delta \left( \frac{1}{2\kappa^2_5} \int d^4x d z
    \sqrt{g} \frac{1}{2} {\cal K}_{\rho\rho}(\partial \rho)^2 \right)
    = \frac{1}{2 \kappa^2_5} \int d^4x \ \delta \rho(\epsilon)
    \left[e^{-3y}{\cal K}_{\rho\rho}\partial_z \rho\right]_{z=\epsilon}\quad.
\end{equation}
On the field theory side, this corresponds to
taking the logarithmic derivative with respect to the square of the gauge
coupling, defined at the ultraviolet cut-off length
$z=\epsilon$
\footnote{In our parametrization of the five-dimensional
    dilaton $\rho\ex{i\alpha}$, the string coupling is given by
    $g_s=\ex{\rho}$ when $\alpha=0$.
}. We get
\begin{equation}
    \langle \int d^4x \frac{1}{g_s} {\rm tr} F^2(x) \rangle =
    \delta^4(0) \frac{\Lambda_\rho}{\kappa^2_5}\quad.
\end{equation}
Taking into account the relation between the asymptotic
five-dimensional gravitational constant and anti-de Sitter
radius $L$
\footnote{Both are values at the boundary.
Remember that the dilaton and the curvature depend
on the bulk coordinate $z$.}
\begin{equation}
\frac{1}{2\kappa_5^2} = \frac{N^2}{2\pi^2 L^3}
\end{equation}
and canceling the four-dimensional space-time volume $\delta^4(0)$,
we get
\begin{equation}
\label{eq:FF}
\langle \frac{1}{g_s} {\rm tr} F^2 \rangle =
\frac{N^2 \Lambda_\rho}{\pi^2 L^3} = \frac{N^2}{16\sqrt{3}\pi^2 L^4}\quad.
\end{equation}

Let us now compute the condensate of the
$SO(3)$-invariant operator
\begin{equation}
{\cal O}_\chi = \frac{1}{\sqrt{3}}\sum_{i=1}^3 {\rm
tr}\lambda^i\lambda^i
\end{equation}
associated to the bulk field $\chi$.
To compute it, we use the formula
\begin{equation}
\label{eq:cond_chi}
\langle \int d^4 x\,{\cal O}_\chi \rangle =
-\left. \frac{\partial}{\partial \chi_0}
\langle e^{-\chi_0 \int {\cal O}_\chi} \rangle \right|_{\chi_0=0}
= \left. -\frac{\partial S_{\rm sugra}(\chi_0)}
{\partial \chi_0}\right|_{\chi_0 =0}
\end{equation}
where $\chi_0$ is a constant source for ${\cal O}_\chi$,
which,  after taking the derivative, we put to zero. That is, we
evaluate it in the supergravity background corresponding
to the $SU(3)$-invariant Lagrangian.
For $\chi_0 \not=0$,
the solution looks like
\begin{equation}
\chi(z) = \chi_0\,z +\Lambda_\chi^3\,z^3 + {\cal O}(\chi_0^2)
\end{equation}
with corrections of ${\cal O}(\chi_0^2)$ to the
remaining functions $\sigma$, $\rho$ and $y$.
Repeating the same manipulations as in the
glueball condensate computation, we get
\begin{equation}
\label{eq:chi_cond}
\langle {\cal O}_\chi \rangle =
\frac{3 N^2}{2\pi^2} \Lambda_\chi^3\quad.
\end{equation}
In the previous section, we observed that, due to the fact that
the infrared solution has $\chi\not=0$, the ultraviolet solution must have
$\Lambda_\chi\not=0$. So the chiral symmetry,
$SU(3)$, really is broken down to $SO(3)$.

Note that  both condensates have the expected field theory
factor of $N^2$, as the fields are in the
adjoint representation of the gauge group. Perhaps surprisingly, the glueball
condensate
\eqref{eq:FF} does not have any explicit $m$-dependence. So long as $m\neq0$,
its value is fixed by \eqref{eq:LambdaRhoDef}. On the other hand, we expect
that $\Lambda_\chi$ is of order $m$.

\subsection{Goldstone Bosons}

The boundary theory has a spontaneously broken global symmetry, so we
expect to see Goldstone bosons.
The spontaneous symmetry breaking of the global symmetry
$SU(3)$ to $SO(3)$ produces 5 Goldstone bosons, which couple to the
broken currents, as reviewed in \S\ref{sec:intro}.
Let us see how their effects arise from the point of view of supergravity.

In the AdS/CFT correspondence, the $SU(4)_R$ currents couple to gauge fields of
the bulk supergravity theory. If the gauge symmetry is unbroken, then the
two-point function of the currents falls off as $1/|x|^6$. For the broken
currents, however, we will find a slower falloff, $1/|x|^4$, as the distance,
$|x|$, between the two insertions goes to infinity. This is the
position-space version of \eqref{eq:two-point},
\begin{equation}
\label{eq:JJ}
\lim_{|x|\to \infty}
\langle J_\mu{}^a(x) J_{\nu}{}^b(0) \rangle =
\delta^{ab} f_\pi^2\,\partial_\mu \partial_\nu
\left(\frac{1}{|x|^2}\right)
+{\cal O}\left(\frac{1}{|x|^6}\right)\quad.
\end{equation}
Our goal will be to extract the pion decay constant, $f_\pi$.
Before we can do that, however, we need to normalize the currents correctly.
To do this, we will choose them to be normalized so as to reproduce the
anomaly \eqref{eq:three-point}. Happily, this computation has already been
done (in the $SU(4)$-invariant AdS theory) in \cite{Freedman:1998tz}. The
anomaly is reproduced by the Chern-Simons term of the
five-dimensional gauged supergravity action. It is unaffected by the
perturbation we are doing.

Returning to the two-point function, we need the quadratic part of the action
evaluated in the $SO(3)$-invariant background.
For the gauge fields not in the unbroken
$SO(3)$ group, there is a mass term in the action generated by
the nonzero profile of the fields $\sigma$ and $\chi$.
In the unitary gauge corresponding to the background solution
\eqref{eq:uvsol} and \eqref{eq:irsol}, the quadratic action
for those gauge fields is
\begin{equation}
\label{eq:actionvec}
S = \frac{1}{2\kappa_5^2} \int dz\,d^4x \,\sqrt{g}
\left(\frac{1}{2e^2} g^{ij}g^{kl}F_{ik}^a F_{jl}^a
+\frac{1}{2}\left({\cal M}^2\right)_{ab}\,g^{ij}A_i{}^a A_j{}^b
\right)
\end{equation}
where $e^2 = 4/L^2$ is the five dimensional $SU(4)$ gauge coupling,
$F_{ij}=\partial_i A_j -\partial_j A_i$ ($i,j =1,\dots,5$; $x^5 \equiv
z$).
The mass matrix ${\cal M}^2{}_{ab}$ decomposes in two blocks.
One, corresponding to the gauge fields not in the $SU(3)$
subgroup, depends on  both $\sigma$ and $\chi$,
and the other, for the breaking
of $SU(3)$ to $SO(3)$, depends only on
the field $\chi$. In terms of the supergravity theory, both
have the same effect: they Higgs some gauge symmetries.
But, as we will see  below, those mass terms involving $\sigma$
behave differently near the boundary and do not give rise to Goldstone bosons.

Let us focus on the gauge fields
that point in the directions of the coset space $SU(3)/SO(3)$.
It is easy to see that their mass matrix is proportional
to the identity; hence we will not write the Lie algebra
indices explicitly.
We have to find the boundary-to-bulk propagator,
\begin{equation}
A_i(x,z) = \int d^4 y\, K_{i\mu}(x-y,z) a_\mu(y)
\end{equation}
with $a_\mu$ the source of the four dimensional
global current $J^\mu$. The kernel $K_{i\mu}(x,z)$
has to satisfy the equations of motion coming from the
action \eqref{eq:actionvec} and the boundary conditions
\begin{equation}
\label{eq:Kbc}
\begin{split}
&\lim_{z\to 0} K_{\nu\mu}(x,z) = \delta_{\nu\mu}\,\delta^4(x)
\\
&\lim_{z\to 0} K_{z\mu}(x,z) = 0\quad.
\end{split}
\end{equation}

We don't know the exact expression for the kernel $K_{i\mu}(x,z)$
in our particular curved background because we only have a perturbative
solution \eqref{eq:uvsol} to the supergravity equations. But
we are only interested in the soft pion limit \eqref{eq:JJ}, and so the
perturbative solution suffices, as all we need is an expansion of
$K_{i\mu}$ in powers of $mz$.

The zero$^{th}$ order expression for the kernel,
$K^{(0)}_{i\mu}$ is just the anti-de Sitter solution,
but when $m\neq 0$ the kernel must satisfy
the integrability condition
\begin{equation}
\partial_z K^{(0)}_{z\nu} +\partial_\mu K^{(0)}_{\mu\nu} =
\frac{3}{z} K^{(0)}_{z\nu}
\end{equation}
This gives
\begin{equation}
\begin{split}
  K^{(0)}_{\mu\nu}(x,z)=& \frac{2 z^2}{9\pi^2 |x|^6(z^2+|x|^2)^3}
\left(\delta_{\mu\nu}|x|^2(6|x|^4+3 z^2 |x|^2 +z^4)\right.
\\
&\left. -4x_\mu x_\nu (3 |x|^4 +3 z^2 |x|^2 +z^4) \right)
\\
K^{(0)}_{z\nu}(x,z) = & 0
\end{split}\label{eq:K0def}
\end{equation}
which, as one can check, satisfies \eqref{eq:Kbc}.

Let us now analyze the first order perturbation in $m$.
To do so, we will assume that $\Lambda_\chi \sim m$;
{\it i.e.}, they are of the same order.  Due
to the fact that the mass matrix is
${\cal M}^2{}_{ab} \sim \delta_{ab} \chi^2
\sim \delta_{ab} \Lambda_\chi^6 z^6$ near the boundary $z=0$,
the first order perturbation of the
gauge boson equations of motion comes only from the metric
\begin{equation}
e^{-y(z)} \sim \frac{L}{z} \left(1-\tfrac{1}{9}(mz)^2\right)\quad.
\end{equation}

There is a unique solution that does not have a pole%
\footnote{
In our peculiar gauge choice, there is such a pole in the zero$^{th}$ order
kernel
\eqref{eq:K0def}, but the singular term is pure-gauge in the
five-dimensional sense. $K^{(1)}_{ij}$ is \emph{not} pure-gauge in the
five-dimensional sense, though $K^{(1)}_{\mu\nu}$ is pure-gauge in the
four-dimensional sense.
}
at $x=0$ for $z\not=0$; it is
\begin{equation}
\begin{split}
K^{(1)}_{\mu\nu}(x,z) &= \frac{(mz)^2}{81\pi^2}
\left(\frac{4x_\mu x_\nu- \delta_{\mu\nu}|x|^2}{(z^2+|x|^2)^3}
-\frac{3 z^2 \delta_{\mu\nu}}{(z^2+|x|^2)^3}\right)
\\
K^{(1)}_{z\nu}(x,z) &=0\quad.
\end{split}
\end{equation}
The two-point function is given by
\begin{equation}
\label{eq:JJ2}
\langle J_\mu(x) J_\nu(y) \rangle =
\left.\frac{\delta^2}{\delta a^\mu(x) \delta a^\nu(y)} S[a]
\right|_{a=0}
\end{equation}
where the on-shell action becomes
\begin{equation}
\begin{split}
\label{eq:vec_action_onshell}
S[a] &= \frac{1}{2\kappa_5^2 e^2} \int d^4x\,
\left[e^{-y} A_\mu F_{z\mu}\right]_{z\to 0}
\\
&=\frac{N^2}{8 \pi^2}\int\,d^4x\,d^4y\,a_\mu(x)
\left[\frac{1}{z}\partial_z K_{\mu\nu}(x-y,z)\right]_{z\to 0} a_\nu(y)\quad.
\end{split}
\end{equation}
If we plug in the solution $K_{i\mu} = K^{(0)}_{i\mu}
+K^{(1)}_{i\mu}$,
into (\ref{eq:JJ2}) and take the limit $|x| \to \infty$, we obtain
the expression (\ref{eq:JJ}), with
\begin{equation}
f_\pi^2 = \frac{N^2 m^2}{324\pi^4}\quad.
\end{equation}
Once again, we obtain the expected color factor, $N^2$.

Note the, at first rather surprising, feature that we can read off the
value of $f_{\pi}$ (an infrared property of the field theory) from the
behaviour of the supergravity solution near the boundary.
The solution for the propagator $K_{ij}(x,z)$
has an expansion in powers of $mz$.
Since we are looking for the leading behavior
in $m$, this turns out to involve the leading
behavior in $z$, {\it i.e.}, the behavior near the boundary.
Physically, the bulk mode which dominates the 2-point
function has a $z$-dependent mass which vanishes near the boundary.
For a large separation of points on the boundary,
it prefers to propagate in a ``skin'' of depth
$\sim 1/m$, rather than penetrating deep into the interior.

Finally, we turn to the currents which couple to the explicitly-broken
generators in
$SU(4)/SU(3)$. In this case, the gauge boson mass term depends of
$\sigma$, giving ${\cal M}^2 \sim \sigma^2 \sim (m z)^2$ near
the boundary. The mass term enters into the
analysis of the perturbed equations of motion, and one finds that the
two-point function of these currents  decays exponentially with distance.

\subsection{Wilson Loops}

Let us now discuss confinement in this theory.
The effective five-dimensional ${\cal N}=8$
$SU(4)$ gauged supergravity
contains twelve two-forms $B^{I\alpha}$ ($I=1,...,6$, $\alpha=1,2$)
transforming as a $({\bf6},{\bf2})$ of $SU(4)\times SL(2,\BR)$.
They couple, in an $SL(2,\BZ)$-invariant fashion, to
the $(p,q)$ type IIB strings moving into the
five dimensional space-time.
The strength of this coupling is given by effective string tensions,
$T_{\rm eff}^{(p,q)}$,
which in general are background dependent
\cite{Girardello:1999hj}.
The values of these tensions are given by the square root of the
eigenvalues corresponding to the $({\bf 6,2})$-block
of the $Sp(4)$ invariant metric
\begin{equation}
\label{eq:A}
G_{I\alpha,J\beta}= V_{I\alpha}{}^{ab} V_{J\beta ab}
\end{equation}
that appears in the kinetic term of the two-forms
\begin{equation}
\epsilon_{\alpha\beta} B^{I\alpha}\wedge dB^{I\beta}
+G_{I\alpha,J\beta} B^{I\alpha}\wedge \star B^{J\beta}\quad.
\end{equation}

The eigenvalues of $G_{I\alpha,J\beta}$ are functions
of $\chi$ and $\sigma$; we give their values in the Appendix A.
At the origin of the field space $\{\chi,\sigma\}$,
the eigenvalues are six copies of $\{e^{2\rho},e^{-2\rho}\}$,
corresponding to the fundamental and D1 strings, respectively.
But when $\chi,\sigma \not=0$, they are grouped in $SO(3)$ triplets.
{}From the point of view of the modified Wilson loops of Maldacena,
this is to be expected. Their
definition includes a term with the adjoint
scalars $X^I$ pointing in a particular direction of $S^5$.
The six dimensional vector representation of $SO(6)$
decomposes into ${\bf 3} + {\bf 3}$ under our $SO(3)$ embedding.

The static potential between a $(p,q)$-type dyon and its anti-dyon
is given by the
world-sheet that minimizes
the action of the $(p,q)$ string in our background.
For a separation $r$, we have
\begin{equation}
E_{(p,q)}(r) = \frac{1}{2\pi\alpha'}
\int_{-\frac{r}{2}}^{\frac{r}{2}} dx\, e^{-2y}T_{\rm
eff}^{(p,q)}(z)\,
\sqrt{1 + \left(\frac{dz}{dx}\right)^2}\quad.
\end{equation}
The infrared behavior of the prefactor
\begin{equation}
e^{-2y}T_{\rm eff}^{(p,q)}
\label{eq:factor}
\end{equation}
typically determines whether these dyons are confined.
If it goes to zero, generically the charge is screened (a Higgs phase)
If, however, if it goes to zero
with a very particular power, we have a Coulomb phase
\cite{Minahan:1999yr,Girardello:1999hj}.
By contrast, if the positive quantity
\eqref{eq:factor} does not go to zero, we typically have confinement.
If it diverges in the infrared,  with the absolute minimum
at some point $z=z_{\rm min}$, then
for large enough dyon-antidyon
separation the $(p,q)$ string is able to reach
  $z=z_{\rm min}$.
The most energetically favorable
situation is to remain there, giving a linear confining potential.

Things are more delicate when the minimum of \eqref{eq:factor} is at $z=z_0$.
For large separation, the string connecting the dyons penetrates all the way to
the singularity, where stringy corrections are important.
This is what happens in our case.
When $y\to \infty$, the two triplets of tensions go like%
\footnote{Remember that the dilaton becomes irrelevant in this
region.}
\begin{subequations}
\begin{align}
&T_{\rm eff}^{(1)} \sim \{e^{2\frac{\chi}{\sqrt{3}}},
e^{-2\frac{\chi}{\sqrt{3}}} \}
\\
&T_{\rm eff}^{(2)} \sim \{e^{\frac{\chi}{\sqrt{3}}+\sigma},
e^{-\frac{\chi}{\sqrt{3}}-\sigma} \}\quad.
\end{align}
\end{subequations}
For the $SO(3)$-invariant supergravity background we have found,
the factor \eqref{eq:factor} goes to zero  as $z\to z_0^-$. Hence,
in order to conclude anything about the behaviour of the Wilson loops, we need
to understand the physics in the neighbourhood of the singularity.

\section{Conclusions}

We have studied chiral symmetry breaking
in the context of the AdS/CFT correspondence. We made
  a relevant deformation of the  ${\cal N}=4$
$SU(N)$ Yang Mills theory by the addition of a mass term for one
of the fermions. We found a unique solution for the deformed
supergravity equations of motion, which corresponds to a
nonzero condensate of a quark bilinear in a singlet representation
of the $SO(3)$ subgroup of $SU(3)$. We also computed the
`pion' decay constant $f_\pi$ of the Goldstone bosons
{}from supergravity.

Ideally, one would like to take $m\to\infty$, and thereby study the
``pure" adjoint QCD theory with $N_f=3$. This is possible, in principle,
for \emph{small} 't Hooft coupling. In that case, the scale at which the
adjoint QCD theory becomes strong is far below the scale $m$, at which
$SU(4)$ is broken to $SU(3)$. The supergravity, however, is valid for
\emph{large}  't Hooft coupling and, in this case, the chiral symmetry
breaking scale is not all that different from the scale at which $SU(4)$
is explicitly broken. To really probe the adjoint QCD theory (which is
defined to be \emph{weakly-coupled} at the cutoff scale $m$), we would need
to study the full type IIB string theory in this background.
Alas, that is beyond the scope of the present work.

\section*{Acknowledgements}

We would like to thank
  David Berenstein, Dan Freed, Willy Fischler, Vadim Kaplunovsky
and Steven Weinberg for valuable discussions.

\appendix

\bigskip
\section{Computation of the Supergravity Lagrangian}

The construction of the $D=5$, ${\cal N}=8$ $SU(4)$ gauged
supergravity Lagrangian is explained in \cite{Gunaydin:1986cu},
and recently reviewed in \cite{Distler:1998gb,Freedman:1999gp}.

We are particularly interested in the scalar sector, which involves
$42$ scalars parametrizing the right coset space
$Sp(4)\setminus E_{6(6)}$.
A point in this space is described by a $27 \times 27$ matrix
${\cal V}^{ab}{}_{AB}$. This $27$-bein is in
the ${\bf \overline{27}}$ of $E_6$ (acting on the
$E_6$ indices $A,B=1,\dots,8$ from the right) and ${\bf 27}$ of
$Sp(4)$ (acting on the $Sp(4)$ indices $a,b=1,\dots,8$ from the left).
It maps the ${\bf 27}$ of $E_{6(6)}$ to the
${\bf 27}$ of $Sp(4)$.

To parametrize the
scalar manifold,  it is sometimes more
convenient to work in a different basis
for the ${\bf 27}$ of $E_{6(6)}$.
In the original reference \cite{Gunaydin:1986cu},
the scalars were  parametrized in
an $SL(6,\BR) \times SL(2,\BR)$
real basis $\{Z_{IJ}, Z^{I\alpha} \}$
($I,J = 1,\dots,6$ and $\alpha =7,8$ are $SL(6,\BR)$
and $SL(2,\BR)$ indices, respectively) for the ${\bf 27}$
of $E_{6(6)}$, which decomposes into
$({\bf \overline{15},1})\oplus ({\bf 6,2})$.
This is useful because the gauged $SO(6)$ embeds easily as a
subgroup of $SL(6,\BR)$.

In this basis, if we denote a generic element of the coset space
$Sp(4)\setminus E_{6(6)}$ by a real matrix
$U = \exp(\Phi)$, the change of basis that relates $U$
to the $27$-bein (which is linked to basis of the local
$Sp(4)$ symmetries) is performed by $8\times 8$ hermitian
and pure imaginary matrices $\Gamma_I$ that satisfy
the $SO(6)$ Clifford algebra%
\footnote{\cite{Freedman:1999gp} gives a particular
realization of the $SO(6)$ Clifford algebra.}, and
\begin{equation}
\Gamma_0 = \pm i\,\prod_{I=1}^6 \,\Gamma_I\quad.
\label{gamma0}
\end{equation}
The sign is determined by the particular
complex structure chosen on the space on which
$SL(6,\BR)\times SL(2,\BR)$ acts.
The relation between ${\cal V}^{ab}{}_{AB}$
and $U^{AB}{}_{CD}$ is given by
\begin{equation}
{\cal V}^{ab}{}_{AB} =
(\Gamma_{IJ})^{ab}\,U^{IJ}{}_{AB}
+ (\Gamma_{I\alpha})^{ab}\,U^{I\alpha}{}_{AB}
\end{equation}
where one defines
\begin{subequations}
\begin{align}
\Gamma_{IJ} = \frac{1}{2}(\Gamma_I \Gamma_J - \Gamma_J \Gamma_I)
\\
\Gamma_{I\alpha} = (\Gamma_I,i\,\Gamma_I \Gamma_0)\quad.
\end{align}
\end{subequations}

To study the supergravity theory
deformed by $SU(3)$-invariant boundary conditions,
it proves convenient to construct the scalar manifold
in an $SU(3)$-adapted basis. The construction was
explained in \cite{Distler:1998gb}, and we follow it here,
except for a small re-ordering of the $SU(3)$ irreps comprising
the ${\bf 27}$ of $E_{6(6)}$
\footnote{To present a clearer change of basis matrix from
$SL(6,\BR)\times SL(2,\BR)$ to $SU(3)$ basis.}. Here we take
\begin{equation}
\begin{pmatrix}
{\bf 1}_{(0,0)}\\
{\bf 8}_{(0,0)}\\
{\bf 3}_{(4,0)}\\
\overline{\bf 3}_{(-4,0)}\\
{\bf 3}_{(-2,2)}\\
\overline{\bf 3}_{(2,2)}\\
{\bf 3}_{(-2,-2)}\\
\overline{\bf 3}_{(2,-2)}
\label{su3basis}
\end{pmatrix}\quad.
\end{equation}

The three $SO(3)$ singlet complex scalars are introduced by the hermitian
matrices (corresponding to noncompact generators)
\begin{equation}
\Phi_\rho =\begin{pmatrix}
0 & 0 & 0 & 0 & 0 & 0 & 0 & 0 \\
0 & 0 & 0 & 0 & 0 & 0 & 0 & 0 \\
0 & 0 & 0 & 0 & 0 & 0 & 0 & 0 \\
0 & 0 & 0 & 0 & 0 & 0 & 0 & 0 \\
0& 0& 0& 0& 0& 0& \rho\ex{i\alpha}& 0 \\
0 & 0 & 0 & 0 & 0 & 0 & 0 & \rho\ex{i\alpha} \\
0 & 0 & 0 & 0 & \rho\ex{-i\alpha} & 0 & 0 & 0 \\
0& 0& 0& 0& 0& \rho\ex{-i\alpha}& 0& 0
\end{pmatrix},\quad
\Phi_\sigma=\begin{pmatrix}
0 & 0 & 0 & 0 & 0 & 0 & 0 & 0 \\
0 & 0 & 0 & 0 & 0 & 0 & 0 & 0 \\
0 & 0 & 0 & 0 & 0 & 0 & \sigma\ex{-i\gamma} & 0 \\
0 & 0 & 0 & 0 & 0 & \sigma\ex{i\gamma} & 0 & 0 \\
0 & 0 & 0 & 0 & 0 & 0 & 0 & 0 \\
0 & 0 & 0 & \sigma\ex{-i\gamma} & 0 & 0 & 0 & 0 \\
0 & 0 & \sigma\ex{i\gamma} & 0 & 0 & 0 & 0 & 0 \\
0 & 0 & 0 & 0 & 0 & 0 & 0 & 0
\end{pmatrix}
\end{equation}
and
\begin{equation}
\Phi_\chi=\begin{pmatrix}
0 & 0 & 0 & 0 & 0 & 0 & 0 & 0 \\
0 & 0 & 0 & 0 & \sqrt{6}\,s^\curlyvee & 0 & 0 &
\sqrt{6}\,\overline{s}^\curlyvee \\
0 & 0 & 0 & 0 & 0 & s & 0 & 0 \\
0 & 0 & 0 & 0 & 0 & 0 & \overline{s} & 0 \\
0 & \sqrt{6}\,\overline{s}^\curlyvee & 0 & 0 & 0 & 0 & 0 & 0 \\
0 & 0 & \overline{s} & 0 & 0 & 0 & 0 & 0 \\
0 & 0 & 0 & s & 0 & 0 & 0 & 0 \\
0 & \sqrt{6}\,s^\curlyvee & 0 & 0 & 0 & 0 & 0 & 0
\end{pmatrix}
\end{equation}
where the bar means complex conjugation and
${\overline s^\curlyvee}^{ij}{}_k =
\epsilon^{ijl}\,\overline{s}_{lk}$.
For the $SO(3)$ singlet, we have
$s^{ij}= \sqrt{6}i \chi e^{i\beta}\,\delta^{ij}$.

To identify $\rho \ex{i\alpha}$ with the complex
flat direction of the potential, {\it i.e}~as
the five dimensional dilaton and axion, we
parametrize the $SO(3)$-invariant submanifold%
\footnote{Strictly speaking, this $SO(3)$ invariant submanifold
has an additional dimension associated to a real field in the
${\bf 20'}$ of $SO(6)$, which we put to zero in our discussion.}
by
\begin{equation}
{\cal U}(\chi \ex{i\beta},\sigma \ex{i\gamma},\rho \ex{i\alpha}) =
\ex{\Phi_\chi +\Phi_\sigma}\,\ex{\Phi_\rho}\quad.
\end{equation}

To obtain the expressions of the $27$-beins
${\cal V}^{ab}{}_{AB}$, we need the change of basis matrix that
relates the $SU(3)$ basis given in (\ref{su3basis}) to the real
basis $\{Z_{IJ}, Z^{I\alpha} \}$.
To relate them, we should first choose a complex
structure on $\BR^6$ and $\BR^2$. We take, respectively,
\begin{equation}
\begin{split}
z^i &= {1\over \sqrt{2}}(x^i + i x^{i+3}) \quad i=1,2,3
\\
u &= \frac{1}{\sqrt{2}}(x^7 + i x^8)\quad.
\end{split}
\end{equation}
This choice forces us to take the minus sign in \eqref{gamma0}.

Next we order the 27 dimensional real basis by the following
nine triplets
\begin{equation*}
\{Z^k_{(d)},Z^k_{(s)},Z^k_{(a)},Z^k_{(0)},Z^k_{(3)},
Z^{k,7},Z^{k,8},Z^{k+3,7},Z^{k+3,8} \}
\end{equation*}
with the definitions:
\begin{gather*}
Z^k_{(d)} =  Z_{k,k+3}\\
Z^1_{(s)} = {1\over \sqrt{2}}(Z_{26}+Z_{35}) \ ,\
Z^2_{(s)} = {1\over \sqrt{2}}(Z_{34}+Z_{16}) \ ,\
Z^3_{(s)} = {1\over \sqrt{2}}(Z_{15}+Z_{24})\\
Z^1_{(a)} = {1\over \sqrt{2}}(Z_{26}-Z_{35}) \ ,\
Z^2_{(a)} = {1\over \sqrt{2}}(Z_{34}-Z_{16}) \ ,\
Z^3_{(a)} = {1\over \sqrt{2}}(Z_{15}-Z_{24})\\
Z^k_{(0)} = {1\over 2}\epsilon^{kij}Z_{i,j}\\
Z^k_{(3)} = {1\over 2}\epsilon^{kij}Z_{i+3,j+3}\quad.
\end{gather*}
For the block ${\bf 8_{(0,0)}}$ in \eqref{su3basis},
we take the ordering $\{H^1_{(d)},H^2_{(d)},H^k_{(s)},H^k_{(a)} \}$
which, in the $SU(3)$ fundamental representation, acts by
\begin{equation}
H^i{}_j = \begin{pmatrix}
H^1_{(d)}+\frac{H^2_{(d)}}{\sqrt 3} & H^3_{(s)}+iH^3_{(a)}
& H^2_{(s)}-iH^2_{(a)} \cr
H^3_{(s)}-iH^3_{(a)} & -H^1_{(d)}+\frac{H^2_{(d)}}{\sqrt 3} &
H^1_{(s)}+iH^1_{(a)} \cr
H^2_{(s)}+iH^2_{(a)} & H^1_{(s)}-iH^1_{(a)} &
-\frac{2}{\sqrt 3}H^2_{(d)}\quad.
\end{pmatrix}
\end{equation}

Then, the unitary change of basis matrix, from the $SL(6)\times SL(2)$
to the $SU(3)$ basis, is
\begin{equation}
M= \begin{pmatrix}
N & 0 & 0 & 0 & 0 & 0 & 0 & 0 & 0 \cr
0 & 1 & 0 & 0 & 0 & 0 & 0 & 0 & 0 \cr
0 & 0 & 0 & {1\over \sqrt{2}} & {1\over \sqrt{2}} & 0 & 0 & 0 & 0 \cr
0 & 0 & {1\over \sqrt{2}} & {i/ 2} & -{i/ 2} & 0 & 0 & 0 & 0 \cr
0 & 0 & {1\over \sqrt{2}} &-{i/ 2} & {i/ 2} & 0 & 0 & 0 & 0 \cr
0 & 0 & 0 & 0 & 0 & {1\over 2} & {i\over 2} & {i\over 2} & -{1\over
2} \cr
0 & 0 & 0 & 0 & 0 & {1\over 2} & {i\over 2} & -{i\over 2} & {1\over
2} \cr
0 & 0 & 0 & 0 & 0 & {1\over 2} & -{i\over 2} & {i\over 2} & {1\over
2} \cr
0 & 0 & 0 & 0 & 0 & {1\over 2} & -{i\over 2} & -{i\over 2} & -{1\over
2}
\end{pmatrix}
\end{equation}
where
\begin{equation*}
N=\begin{pmatrix}
\frac{1}{\sqrt 3} & \frac{1}{\sqrt 3} & \frac{1}{\sqrt 3} \cr
\frac{1}{\sqrt 2} & -\frac{1}{\sqrt 2} & 0 \cr
\frac{1}{\sqrt 6} & \frac{1}{\sqrt 6} & -\sqrt{\frac{2}{3}}
\end{pmatrix}
\end{equation*}
is the orthogonal transformation that takes the diagonal entries
$Z^k_{(d)}$ to the ${\bf 1}_{(0,0)}$, $H^1_{(d)}$ and
$H^2_{(d)}$.

We finally have all the ingredients to compute the
$SO(3)$-singlet scalar sector of the supergravity Lagrangian.
The kinetic terms are obtained by projecting onto the
non-compact generators of $E_{6(6)}$, which ensures
invariance under the local left-action of $Sp(4)$ on ${\cal U}$. Since
compact generators are realized by
antihermitian matrices, the projection is easily implemented by  the formula
\begin{equation}
{\cal K} =-\frac{1}{48} {\rm tr}\left\{\left({\cal U}\partial{\cal U}^{-1}
+\left({\cal U}\partial{\cal U}^{-1}\right)^\dagger\right)^2\right\}\quad.
\end{equation}
The scalar potential is
\begin{equation}
V =-\frac{e^2}{32}\left( 2W^{ab}W_{ab} -W^{abcd}W_{abcd}\right)
\end{equation}
with
\begin{equation}
W^{abcd} =\epsilon^{\alpha\beta}\delta^{IJ}{\cal V}^{ab}{}_{I\alpha}
{\cal V}^{cd}{}_{J\beta}\,,\quad W^{ab}=W^c{}_{acb}
\end{equation}
and $SU(4)$ gauge coupling $e=2/L$ in order to have AdS$_5$
space with radius $L$ at the origin of the field space.
Assembling all the pieces, one obtains the Lagrangian written
in \eqref{eq:Lagr}.

As in the tetrad formalism for General Relativity,
one can define an $Sp(4)$ invariant and $E_{6(6)}$
covariant metric form the $27$-beins \cite{Cremmer:1980gs},
\begin{equation}
\label{sp4inv}
G_{AB,CD} = {\cal V}^{ab}{}_{AB} \Omega_{ac}\Omega_{bd}
{\cal V}^{cd}{}_{CD}
\end{equation}
where $\Omega_{ab}$ is the $Sp(4)$ symplectic metric.
Since there is no quadratic invariant for $E_{6(6)}$,
this metric should be used to construct the terms in the
supergravity Lagrangian that involve the vector and antisymmetric
fields (except for the topological Chern-Simons terms).
See \cite{Gunaydin:1986cu} for the expressions of these terms.

To compute the Wilson loops from the
five-dimensional supergravity theory, we will need the eigenvalues
of the $({\bf 6,2})$-block $G_{I\alpha,J\beta}$.
In the $SO(3)$-invariant background they split in two different
$SO(3)$ triplets. One is
\begin{equation}
\begin{split}
&\left.T^{(1)}_{eff}\right.^2 = \frac{1}{2}\,\cosh(2\rho)\,
\left(\cosh\left(\frac{4\chi}{\sqrt 3}\right)
+\cosh\left(\frac{2\chi}{\sqrt 3}-2\sigma\right)\right)
\\
&\pm \frac{1}{2}\sqrt{
\textstyle \sinh^2(2\rho)\,
\left(\cosh\left(\frac{4\chi}{\sqrt 3}\right)
+\cosh\left(\frac{2\chi}{\sqrt 3}-2\sigma\right)\right)^2
+\left(\cosh\left(\frac{4\chi}{\sqrt 3}\right)
-\cosh\left(\frac{2\chi}{\sqrt 3}-2\sigma\right)\right)^2}
\end{split}
\end{equation}
and the other is
\begin{equation}
\left.T^{(2)}_{eff}\right.^2 = \cosh(2\rho)\,
\cosh^2\left(\frac{\chi}{\sqrt 3}+\sigma\right)
\pm \sqrt{\sinh^2(2\rho)
\,\cosh^4\left(\frac{\chi}{\sqrt 3}+\sigma\right)
+\sinh^4\left(\frac{\chi}{\sqrt 3}+\sigma\right)}\quad.
\end{equation}

Finally, we correct Table 1 of \cite{Distler:1998gb}
for the scaling dimensions
of the spinless conformal operators at the $SU(3)$ critical point.
Observe that the scaling dimension of the operator
${\cal O}^{ij}={\rm tr}\lambda^i\lambda^j$ is complex, indicating
that it violates the Breitenlohner-Freedman bound.
\begin{center}
\begin{tabular}{|c|c|c|c|c|c|}
\hline
CFT operator&Supergravity field&$SU(3)\times
U(1)$&$\Delta_{UV}$&$\Delta_{IR}$\\
\hline\hline
$\abs{\mathrm{tr}(\lambda^4 \lambda^4)} $&$\sigma$&
${\bf 1}_{0}$  & $3$ & $2 + 2\sqrt{3} = 5.4641 \cdots $\\
$\mathrm{tr}(\lambda^i \lambda^j) $&$s^{ij}$&
${\bf 6}_{8}$  & $3$ & $2 +i\frac{2}{3} $\\
\hline
$\mathrm{tr}(Z^i Z^j) $&$t^{ij}$&
${\bf 6}_{-4}$  & $2$ & $2+\frac{2}{5}\sqrt{5} =2.8944\cdots $\\
$\mathrm{tr}(
Z^{i}\overline{Z}_{j}-\tfrac{\delta^{i}{}_{j}}{3}
\overline{\boldsymbol{Z}}\boldsymbol{Z})
$&$h^i{}_j$&
${\bf 8}_{0}$  & $2$ & $4$ \\
\hline
\end{tabular}
\end{center}


\section{The Cohomology of $SU(3)/SO(3)$}

The Wess-Zumino Lagrangian is normalized to be the generator of
$\coho{5}{SU(3)/SO(3),\BZ}$. We would like to write this in terms of its
pullback to $SU(3)$. To do that, we use the Leray Spectral sequence
(see \cite{Bott-Tu}, p.~169)
for the fiber bundle $SU(3)\stackrel{\pi}{\to}SU(3)/SO(3)$
\footnote{We would like to thank Dan Freed for helping us through this
computation.}.

The cohomology ring of $SU(3)$ is torsion-free, with generators in
dimensions 3 and 5. That is,
\begin{center}
     \begin{tabular}{|c||c|c|c|c|}
         \hline
         Cohomology& $\coho{0}{SU(3),\BZ}$ &
         $\coho{3}{SU(3),\BZ}$ & $\coho{5}{SU(3),\BZ}$
         & $\coho{8}{SU(3),\BZ}$  \\
         \cline{2-5}
	group & $\BZ$ & $\BZ$ & $\BZ$ & $\BZ$  \\
         \hline
         generator  & 1 & $x_{3}$ & $x_{5}$ & $x_{3}x_{5}$ \\
         \hline
     \end{tabular}
\end{center}
The cohomology of $SO(3)$ has torsion in ${\rm H}^{2}$. The generator
$a$ satisfies $2a=0$.
Note that $a$ is dual to the nontrivial 1-cycle in ${\rm
H}_{1}(SO(3),\BZ)=\BZ_{2}$ which originates because $\pi_{1}(SO(3))=\BZ_{2}$.
\begin{center}
     \begin{tabular}{|c||c|c|c|}
         \hline
         Cohomology & $\coho{0}{SO(3),\BZ}$ &
         $\coho{2}{SO(3),\BZ}$ & $\coho{3}{SO(3),\BZ}$ \\
         \cline{2-4}
	group & $\BZ$ &
         $\BZ_{2}$ & $\BZ$ \\
         \hline
         generator  & 1 & $a$ & $x$  \\
         \hline
     \end{tabular}
\end{center}
{}From this data, we can compute the cohomology of $X=SU(3)/SO(3)$
(along with the information we are after) from the spectral sequence. To
simplify the discussion, we just state the result for the cohomology of $X$,
bearing in mind that it can be \emph{derived} from the analysis below.
\begin{center}
     \begin{tabular}{|c||c|c|c|}
         \hline
         Cohomology & $\coho{0}{X,\BZ}$ & $\coho{3}{X,\BZ}$ &
         $\coho{5}{X,\BZ}$  \\
         \cline{2-4}
         group & $\BZ$ & $\BZ_{2}$ & $\BZ$  \\
         \hline
         generator & 1 & $b$ & $y$  \\
         \hline
     \end{tabular}
\end{center}
Again, the torsion element $b$, with $2b=0$, arises because ${\rm
H}_{2}(X,\BZ)=\pi_{2}(X)=\pi_{1}(SO(3))=\BZ_{2}$.

The Leray Spectral Sequence has as its $E_{2}$ term,
$E_{2}^{p,q}=\coho{p}{X,\coho{q}{SO(3)}}$. These groups can be
computed from the Universal Coefficients Theorem (\cite{Bott-Tu}, p.~194),
\begin{equation*}
     \coho{p}{X,A}={\rm Hom}({\rm H}_{p}(X),A)\oplus{\rm Ext}({\rm
     H_{p-1}}(X),A)
\end{equation*}
where
\begin{equation*}
     \begin{aligned}
          {\rm Ext}(\BZ_{m},\BZ) &=\BZ_{m}   \\
           {\rm Ext}(\BZ_{m},\BZ_{n})& =\BZ_{gcd(m,n)}  \\
         {\rm Ext}(\BZ,\cdot) & =0\quad.
     \end{aligned}
\end{equation*}
The non-obvious cohomology group that one obtains is
$\coho{2}{X,\coho{2}{SO(3)}}={\rm Hom}(\BZ_{2},\BZ_{2})=\BZ_{2}$. We
call its generator $c$.

There is the differential $d_{r}:E_{r}^{p,q}\to E_{r}^{p-r+1,q+r}$, and to
obtain the $E_{r+1}$ term of the spectral sequence, we are instructed
to take the cohomology of $d_{r}$.
\setlength{\unitlength}{1ex}
\begin{center} $E_{2}^{p,q}$
     \begin{tabular}{c|cccccc}
         3 & $x$ &  &  & $xb$ &  & $xy$  \\
         2 & $a$ &
\begin{picture}(2,2)\put(-2,3){\vector(3,-1){6}
\makebox(-6,0){$\scriptstyle d_2$}}\end{picture} &
$c$ &
$ab$ &
\begin{picture}(2,2)\put(-2,3){\vector(3,-1){6}
\makebox(-6,0){$\scriptstyle d_2$}}\end{picture}& $ay$  \\
         1 &  &  &  &  &  &   \\
         0 & $1$ &  &  & $b$ &  & $y$  \\
         \hline
          & 0 & 1 & 2 & 3 & 4 & 5  \\
     \end{tabular}
\end{center}
For the $E_{2}$ term, we see that $d_{2}(xb)=ay$ and $d_{2}(x)=c$. Now,
both $xb$ and $ay$ are of order 2. So, when we take the cohomology,
they both get killed. However, $x$ has infinite order, while $2c=0$. So
$d_{2}(2x)=0$, and hence $2x$ survives in the cohomology.
\begin{center} $E_{3}^{p,q}$
     \begin{tabular}{c|cccccc}
         3 & $2x$ &  &  &  &  & $xy$  \\
         2 & $a$ &  &  & $ab$ &  &   \\
         1 & &\multicolumn{2}{c}{
\begin{picture}(2,2)\put(-4,3.5){\vector(3,-2){9}
\makebox(-10,-3){$\scriptstyle d_3$}}\end{picture}}
  &  &  &   \\
         0 & $1$ &  &  & $b$ &  & $y$  \\
         \hline
          & 0 & 1 & 2 & 3 & 4 & 5  \\
     \end{tabular}
\end{center}
For the $E_{3}$ term, $d_{3}(a)=b$, and the spectral sequence
converges at the $E_{4}$ term.
\begin{center} $E_{4}^{p,q}=E_{\infty}^{p,q}$
     \begin{tabular}{c|cccccc}
         3 & $2x$ &  &  &  &  & $xy$  \\
         2 &  &  &  & $ab$ &  &   \\
         1 & & & &  &  &   \\
         0 & $1$ &  &  &  &  & $y$  \\
         \hline
          & 0 & 1 & 2 & 3 & 4 & 5  \\
     \end{tabular}
\end{center}
Finally, we have
\begin{equation*}
     \begin{aligned}
         G^{(n)} & =\bigoplus_{p+q=n} E_{\infty}^{p,q}  \\
          & = Gr(\coho{n}{SU(3)})\quad.
     \end{aligned}
\end{equation*}
That is, $\coho{n}{SU(3)}$ has a filtration
\begin{equation*}
     \coho{n}{SU(3)}=F_0\supset F_1\supset F_2\supset F_3\supset\dots
\end{equation*}
such that $F_i/F_{i+1}=E^{i,n-i}_\infty$. The nontrivial case here is
$\coho{5}{SU(3)}=\BZ$. We have $F_2/F_3=\BZ_2$ and $F_5/F_6=\BZ$.
This means that $F_i=0$ for $i\geq6$ and $F_i=\BZ$ for $i=0,\dots 5$.
The inclusions $F_i\hookrightarrow F_{i-1}$ for $i<6$ are all isomorphisms,
{\it except} for the map $F_3\hookrightarrow F_2$, which is multiplication by
two. The net effect is that the map
\begin{equation*}
     \pi^*: \coho{5}{SU(3)/SO(3),\BZ}\to\coho{5}{SU(3),\BZ}
\end{equation*}
is multiplication by \emph{two}.

This factor of two resolves an apparent puzzle \cite{DHoker-Weinberg:G/H}.
Since $\pi_{4}(SO(3))=\BZ_{2}$,
there is an ambiguity in lifting the $\sigma$-model map
into $SU(3)/SO(3)$ to a map into $SU(3)$. Let $w$ be the generator
of $\pi_{4}(SO(3))$.
Then both $U(x)$ and $\tilde U(x)=w(x)U(x)$ represent
the {\it same} sigma model
map into the coset space $SU(3)/SO(3)$. In particular, we must have
$e^{-S_{WZ}[\tilde U]}=e^{-S_{WZ}[U]}$. With the naive
normalization of $S_{WZ}$
coming from the generator of $\coho{5}{SU(3),\BZ}$, we would instead find
\begin{equation}
        e^{-S_{WZ}[w]}=-1\quad.
        \label{eq:GlobalAnom}
\end{equation}
But, from the previous analysis,
we find that the correct normalization of $S_{WZ}$,
coming from the generator of $\coho{5}{SU(3)/SO(3),\BZ}$, is {\it twice}
the naive normalization, and thus $e^{-S_{WZ}[w]}=1$, as required.

Similar computations can be carried out in the case of four flavours,
where $\pi_{4}(SO(4))=\BZ_{2}+\BZ_{2}$ and five flavours,
where $\pi_{4}(SO(5))=\BZ_{2}$. Indeed, the sign
in \eqref{eq:GlobalAnom} is intimately related
to the global anomaly \cite{Witten:1982fp}
in gauging $H$. So long as the underlying fermion
theory is anomaly-free, the normalization of $S_{WZ}$
must be such as to eliminate this sign.

\renewcommand{\baselinestretch}{1} \normalsize
\bibliography{so3}
\bibliographystyle{utphys}

\end{document}